\documentclass[12pt,preprint]{aastex}

\slugcomment{%
To appear in the Astrophysical Journal Letters%
\\ %
{\it Submitted 27 April 2007%
; Accepted 14 May 2007%
}}

\shorttitle{T\lowercase{r}ES-3} 
\shortauthors{O'Donovan et al.}

\newcommand{\tresOne}{\mbox{\object[NAME TrES-1]{TrES-1}}}
\newcommand{\tresTwo}{\mbox{\object[GSC 03549-02811]{TrES-2}}}

\newcommand{\hdOEN}{\mbox{\object{HD 189733b}}}
\newcommand{\ogletrFSb}{\mbox{\object{OGLE-TR-56b}}}
\newcommand{\ogletrOOTb}{\mbox{\object{OGLE-TR-113b}}}
\newcommand{\ogletrOTTb}{\mbox{\object{OGLE-TR-132b}}}

\newcommand{\wasptwob}{\mbox{\object[GSC 00522-01199]{WASP-2b}}}
\newcommand{\tresThreeS}{\mbox{\object[GSC 03089-00929]{TrES-3}}}
\newcommand{\tresThreeP}{\mbox{TrES-3}}
\newcommand{\gscTresThree}{\mbox{\object{GSC 03089-00929}}}
\newcommand{\thetaher}{\mbox{\object[HD 163770]{$\theta$ Her}}}

\def\kms{\ifmmode{\rm km\thinspace s^{-1}}\else km\thinspace s$^{-1}$\fi}
\def\ms{\ifmmode{\rm m\thinspace s^{-1}}\else m\thinspace s$^{-1}$\fi}
\newcommand\gsim{\mathrel{\rlap{\lower4pt\hbox{\hskip1pt$\sim$}}\raise1pt\hbox{$>$}}}
\newcommand\lsim{\mathrel{\rlap{\lower4pt\hbox{\hskip1pt$\sim$}}\raise1pt\hbox{$<$}}}

\begin{document}

\title{T\lowercase{r}ES-3: A Nearby, Massive, Transiting Hot Jupiter in a 31-Hour Orbit\altaffilmark{1} }

\author{%
Francis~T.~O'Donovan\altaffilmark{2},
David~Charbonneau\altaffilmark{3, 4},
G\'{a}sp\'{a}r~\'{A}.~Bakos\altaffilmark{3,5}, 
Georgi~Mandushev\altaffilmark{6}, 
Edward~W.~Dunham\altaffilmark{6},
Timothy~M.~Brown\altaffilmark{7,8},
David~W.~Latham\altaffilmark{3}, 
Guillermo~Torres\altaffilmark{3},
Alessandro~Sozzetti\altaffilmark{3,9},
G\'{e}za~Kov\'{a}cs\altaffilmark{10},
Mark~E.~Everett\altaffilmark{11}, 
Nairn~Baliber\altaffilmark{7,8}, 
M\'{a}rton~G.~Hidas\altaffilmark{7,8},
Gilbert~A.~Esquerdo\altaffilmark{3,11},
Markus~Rabus\altaffilmark{12},
Hans~J.~Deeg\altaffilmark{12},
Juan~A.~Belmonte\altaffilmark{12}, 
Lynne~A.~Hillenbrand\altaffilmark{2},
Robert~P.~Stefanik\altaffilmark{3}%
}

\altaffiltext{1}{Some of the data presented herein were obtained at the W.M. Keck Observatory from telescope time allocated to NASA through the agency's scientific partnership with Caltech and the University of California.
The Observatory was made possible by the generous financial support of the W.M. Keck Foundation.}

\altaffiltext{2}{California Institute of Technology, 1200 E.~California Blvd., Pasadena, CA 91125; ftod@caltech.edu}

\altaffiltext{3}{Harvard-Smithsonian Center for Astrophysics, 60 Garden St., Cambridge, MA 02138}

\altaffiltext{4}{Alfred P.~Sloan Research Fellow}

\altaffiltext{5}{Hubble Fellow}

\altaffiltext{6}{Lowell Observatory, 1400 West Mars Hill Rd., Flagstaff, AZ 86001}

\altaffiltext{7}{Las Cumbres Observatory Global Telescope, 6740 Cortona Dr.~Ste.~102, Goleta, CA 93117}

\altaffiltext{8}{Department of Physics, University of California, Santa Barbara, CA 93106}

\altaffiltext{9}{INAF-Osservatorio Astronomico di Torino, 10025 Pino Torinese, Italy}

\altaffiltext{10}{Konkoly Observatory, P.O.~Box 67, H-1525, Budapest, Hungary}

\altaffiltext{11}{Planetary Science Institute, 1700 East Fort Lowell Rd.~Ste.~106, Tucson, AZ 85719}

\altaffiltext{12}{Instituto de Astrof{\'\i}sica de Canarias, 38200 La Laguna, Tenerife, Spain}

\begin{abstract}

We describe the discovery of a massive transiting hot Jupiter with a very short orbital period (\mbox{$1.30619$\,d}), which we name \tresThreeP.
From spectroscopy of the host star \gscTresThree, we measure \mbox{$T_{\rm eff} = 5720 \pm 150$\,K}, \mbox{$\log{g}=4.6 \pm 0.3$}, and \mbox{$v\sin{i} < 2\,\kms$}, and derive a stellar mass of \mbox{$0.90\pm0.15\,M_{\sun}$}.
We estimate a planetary mass of \mbox{$1.92 \pm 0.23\,M_{\rm Jup}$}, based on the sinusoidal variation of our high-precision radial velocity measurements.
This variation has a period and phase consistent with our transit photometry. 
Our spectra show no evidence of line bisector variations that would indicate a blended eclipsing binary star.
From detailed modeling of our $B$ and $z$ photometry of the 2.5\%-deep transits, we determine a stellar radius \mbox{$0.802\pm0.046\,R_{\sun}$} and a planetary radius \mbox{$1.295\pm0.081\,R_{\rm Jup}$}. 
\tresThreeP\ has one of the shortest orbital periods of the known transiting exoplanets, facilitating studies of orbital decay and mass loss due to evaporation, and making it an excellent target for future studies of infrared emission and reflected starlight.

\end{abstract}


\section{Introduction}
\label{sec:intro}

The OGLE-III deep-field survey \citep[][and references therein]{Udalski_Pietrzynski_Szymanski:acta:2003a} has identified 3 transiting planetary systems with very short orbital periods: \ogletrFSb\ ($P=29$\,hr; \citealt{Konacki_Torres_Jha:nat:2003a}), \ogletrOOTb\ ($P=34$\,hr; \citealt{Bouchy_Pont_Santos:aa:2004a, Konacki_Torres_Sasselov:apjl:2004a}), and \ogletrOTTb\ ($P=41$\,hr; \citealt{Bouchy_Pont_Santos:aa:2004a}). 
At the time of discovery, such 1-d period planets were conspicuously absent from radial velocity (RV) surveys. 
These surveys had not discovered a planet with an orbital period of substantially less than 3\,d, despite the sensitivity of such surveys to planets with short orbital periods.
Several authors reconciled this apparent discrepancy by determining the respective observational biases and selection effects \citep{Gaudi_Seager_Mallen-Ornelas:apj:2005a, Pont_Bouchy_Melo:aa:2005a, Gould_Dorsher_Gaudi:acta:2006a, Fressin_Guillot_Morello:preprint:2007a}. 
Since then, hot Jupiters with periods just over 2\,d have been detected by both RV surveys (\hdOEN; \citealt{Bouchy_Udry_Mayor:aa:2005a}) and wide-field transit surveys (\wasptwob; \citealp{Collier-Cameron_Bouchy_Hebrard:MNRAS:2007a}), indicating a consistent picture for the underlying distribution of orbital periods of hot Jupiters. 

With the discovery of these nearby planets, we continue to observe \citep{Sozzetti_Torres_Charbonneau:preprint:2007a} the decreasing linear relation between $P$ and the planetary mass ($M_{p}$) for transiting planets  first noticed by \cite{Mazeh_Zucker_Pont:mnras:2005a} and \cite{Gaudi_Seager_Mallen-Ornelas:apj:2005a}.
It appears from this small number of planets that shorter period exoplanets are on average more massive.
This may suggest different evolution of these planetary systems than the canonical migration (\citealp[see, e.g.,][]{Trilling_Benz_Guillot:apj:1998a}) of hot Jupiters from beyond the ice line.

We present here the discovery of \tresThreeP, a massive planet which has one of the shortest periods of the transiting planets, and which was independently identified by two transit surveys. 

\section{Observations and Analysis}
\label{sec:observations}

Transits of the parent star \gscTresThree\ were detected by two 10-cm telescopes, part of the Trans-atlantic Exoplanet Survey (TrES) network.
Sleuth (Palomar Observatory, California) observed 7 full transits and 3 partial transits, of which 3 full and 2 partial transits were observed by the Planet Search Survey Telescope (Lowell Observatory, Arizona; \citealt{Dunham_Mandushev_Taylor:pasp:2004a}).
We obtained these observations between UT 2006 May 6 and 2006 August 2, during which time we monitored a \mbox{$5\fdg7\times5\fdg7$} field of view (FOV) centered on \thetaher.
We then processed the data from each telescope separately, as we have described previously \citep{Dunham_Mandushev_Taylor:pasp:2004a, ODonovan_Charbonneau_Torres:apj:2006a, ODonovan_Charbonneau_Alonso:preprint:2007a}.
We then search our binned light curves using the box-fitting transit search algorithm of \cite{Kovacs_Zucker_Mazeh:aa:2002a} for periodic event consistent with the passage of a Jupiter-sized planet across a solar-like star.
We flagged \tresThreeP\ as a candidate, noting that the transit duration of 1.3\,hr implied a high impact parameter $b$.

Transits of \tresThreeS\ were independently observed by the Hungarian Automated Telescope Network (HATNet; \citealt{Bakos_Noyes_Kovacs:pasp:2004a}).
We observed the HAT field ``G196'' between UT 2005 June 8 to December 5, using the HAT-7 telescope at the Fred~L.~Whipple Observatory (FLWO) and HAT-9 at the Submillimeter Array atop Mauna Kea.
We applied standard calibration and aperture photometry procedures to the frames as described earlier in \cite{Bakos_Noyes_Kovacs:apj:2007a}.
We also applied the trend filtering algorithm \citep{Kovacs_Bakos_Noyes:mnras:2005a}, and identified \tresThreeP\ as a transit candidate using the algorithm of \cite{Kovacs_Zucker_Mazeh:aa:2002a}. 

We observed \tresThreeS\ using the CfA Digital Speedometers \citep{Latham:ASP:1992a} on 13 occasions from 2006 September 9 to 2007 April 6.
These spectra are centered at 5187\,\AA\ and cover 45\,\AA\ with a resolving power of \mbox{$\lambda/\Delta\lambda \approx 35,\!000$}.
We derived RVs at each epoch by cross-correlation against synthetic spectra created by J.~Morse using Kurucz model stellar atmospheres (J.~Morse \& R.~L.~Kurucz~2004, private communication).
There is no significant variation in these measurements, which have a mean of \mbox{+9.58\,\kms} and a scatter of \mbox{0.73\,\kms}.
This limits the mass of the companion to be less than about $7\,M_{\rm Jup}$.
We estimated the stellar parameters from a cross-correlation analysis similar to that above, assuming a solar metallicity.
The effective temperature ($T_{\rm eff}$), surface gravity ($\log g$), and projected rotational velocity ($v \sin i$) we derive are listed in Table~\ref{tab:tres3}.
Based on these values, we estimate the stellar mass to be $M_{\star}=0.90\pm0.15\,M_{\sun}$.
Further spectroscopic analysis of the parent star (similar to that performed for \tresOne\ and \tresTwo; \citealt{Sozzetti_Yong_Torres:apjl:2004a, Sozzetti_Torres_Charbonneau:preprint:2007a}) is in progress and will be presented elsewhere.

The 2MASS $J-K$ color (0.407\,mag) and UCAC2 proper motion (\mbox{$39.1\,\mathrm{mas\,yr^{-1}}$}) of \gscTresThree\ are as expected for a nearby G-dwarf. 
We obtained absolute $BVR_{C}I_{C}$ photometry for \tresThreeS\ on the night of UT 2007 April 14 with the 105-cm Hall telescope at Lowell Observatory.
The photometry was calibrated using seven standard fields from \cite{Landolt:aj:1992a}. 
The results are listed in Table~\ref{tab:tres3}.

We observed a full transit in the $z$-band on UT 2007 March 26 using KeplerCam \citep[see, e.g.,][]{Holman_Winn_Latham:apj:2006a} at the \mbox{FLWO 1.2-m} telescope. 
We obtained 150 90-s exposures of a $23\farcm1 \times 23\farcm1$ FOV containing \tresThreeS. 
We observed another transit in Bessell $B$ on UT 2007 April 8 with the Las Cumbres Observatory Global Telescope 2-m Faulkes Telescope North on Haleakala, Maui, Hawaii. 
The CCD camera imaged a $4.6\arcmin$ square field with an effective pixel size of $0\farcs28$ (the images were binned $2\times2$). 
We obtained 126 images with an exposure time of 60\,s and a cadence of 70\,s.
We defocused the telescope slightly (to an image diameter of about $3\arcsec$) to avoid saturation of the brightest comparison star.
We analyzed these data sets separately.
Using standard IRAF procedures, we corrected the images for bias, dark current, and flat-field response.
We determined fluxes of the target star and numerous comparison stars using synthetic aperture photometry.
To correct the target star fluxes for time-varying atmospheric extinction, we divided them by a weighted average of the fluxes of all the comparison stars.

We collected high-precision RV measurements on UT 2007 March 27--29 using HIRES \citep{Vogt_Allen_Bigelow:SPIE:1994a} and its I$_2$ absorption cell on the Keck~I telescope. 
We obtained a total of 7 star+iodine exposures and 1 template spectrum.
All spectra were gathered with a nominal resolving power \mbox{$\lambda/\Delta\lambda\simeq 55\,000$}, using the HIRES setup with the $0.86\arcsec$ slit.
We used an integration time of 15\,min, which yielded a typical signal-to-noise ratio of \mbox{$110\,\mathrm{pixel}^{-1}$}. 
We reduced the raw spectra using the MAKEE software written by T.~Barlow. 
We have described the software used to derive relative radial velocities with an iodine cell in earlier works \citep[e.g.,][]{ODonovan_Charbonneau_Mandushev:apjl:2006a, Sozzetti_Torres_Latham:apj:2006a}.
We estimate our internal errors from the scatter about the mean for each spectral order divided by the square root of the number of orders containing I$_2$ lines, and find them to be around $10\,\ms$.
The radial-velocity measurements are listed in Table~\ref{tab:rvtres3}.

\begin{figure}
\plotone{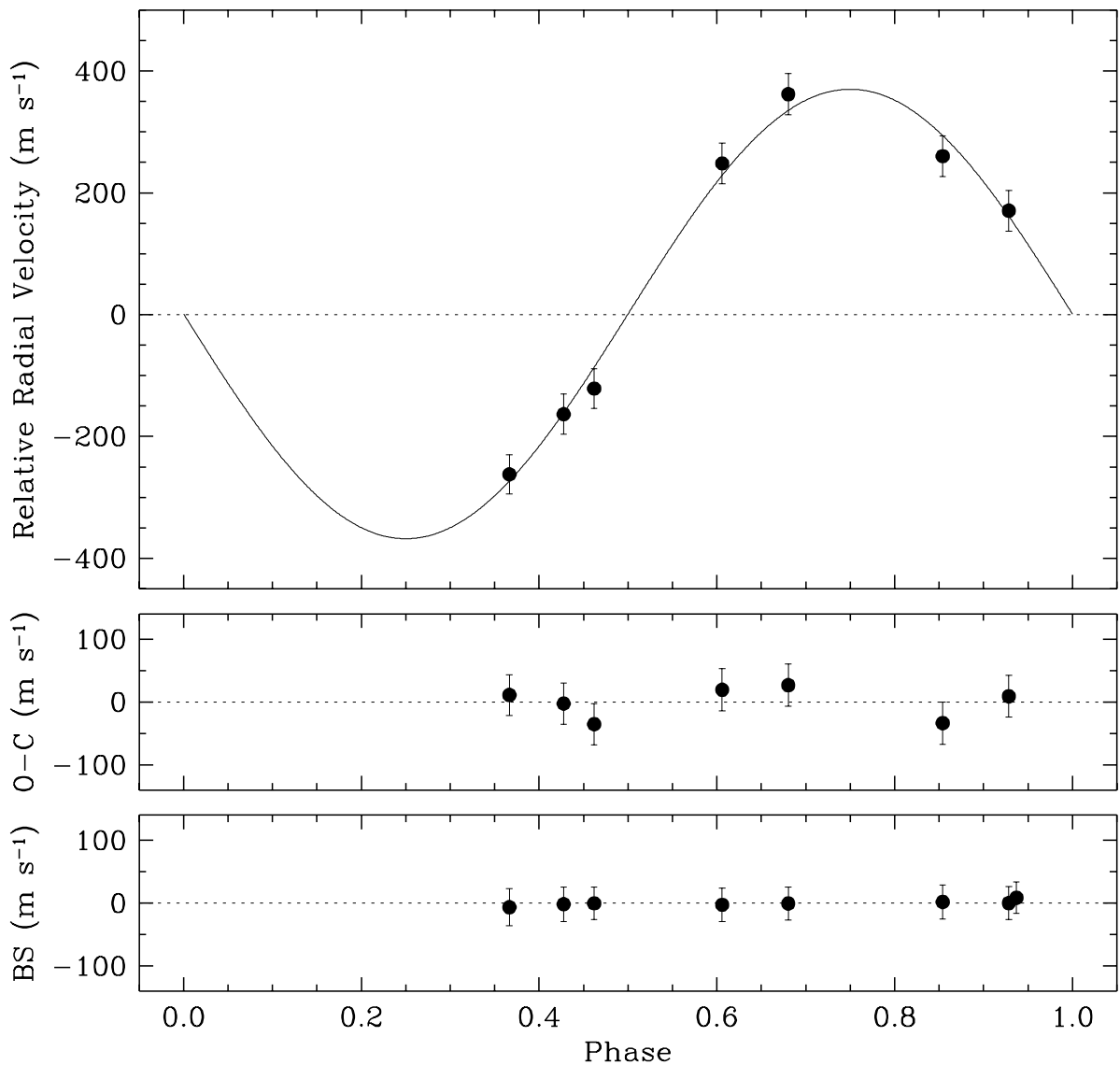}
\caption{{\it Top}: Radial-velocity observations of TrES-3 obtained with Keck/HIRES using the $\mathrm{I}_2$ cell, shown relative to the center of mass and adopting the ephemeris in Table~\ref{tab:tres3b}. 
The best-fit orbit ({\it solid line}) is overplotted.
{\it Middle}: Residuals from the best-fit model to the radial velocities.
{\it Bottom}: Bisector spans shifted to a median of zero, for each of the iodine exposures as well as for the template (which is shown as the additional data point at phase 0.937).%
}
\label{fig:rvtres3}
\end{figure}

We constrained the orbital fit to these data to have zero eccentricity ($e$), as expected from theoretical arguments for such a short period planet, and we also held $P$ and the transit epoch $T_c$ fixed at their values determined from the photometric data. 
The rms residual from this fit (15.4\,\ms) is larger than the typical internal errors (10\,\ms).
Preliminary analysis of the template spectrum suggests that the star shows evidence of activity.
For the inferred spectral type, the presence of radial-velocity ``jitter'' of \mbox{10--20\,\ms} is not unexpected \citep{Saar_Butler_Marcy:apjl:1998a, Santos_Mayor_Naef:aa:2000a, Wright:pasp:2005a}, and would explain the excess scatter we find.
Figure~\ref{fig:rvtres3} displays the RV data overplotted with the best-fit model, along with the residuals from the fit.
The parameters of the orbital solution are listed in Table~\ref{tab:tres3b}.
We find a minimum planetary mass of \mbox{$M_p \sin i = 2.035 \pm 0.090 [(M_p+M_{\star})/M_{\sun}]^{2/3} M_{\rm Jup}$}, where $i$ represents the orbital inclination. 
As a consistency check we repeated the fit with $P$ fixed and $e = 0$ as before, but
solving for $T_c$.
The result is \mbox{$T_c = \mathrm{HJD}\,2,\!454,\!185.911 \pm 0.045$}, which is consistent with, but less precise than, the value determined from the photometry (Table~\ref{tab:tres3b}).

We investigated the possibility that the velocity variations are due not to a planetary companion, but instead to distortions in the line profiles caused by an unresolved eclipsing binary \citep{Santos_Mayor_Naef:aa:2002a, Torres_Konacki_Sasselov:apj:2005a}. 
We cross-correlated each spectrum against a synthetic template matching the properties of the star, and averaged the correlation functions over all orders blueward of the region affected by the iodine lines.
From this representation of the average spectral line profile we computed the mean bisectors, and as a measure of the line asymmetry we calculated the ``bisector spans'' as the velocity difference between points selected near the top and bottom of the mean bisectors \citep{Torres_Konacki_Sasselov:apj:2005a}.
If the velocity variations were the result of a stellar blend, we would expect the bisector spans to vary in phase with the photometric period with an amplitude similar to that seen in the RVs \citep{Queloz_Henry_Sivan:aa:2001a, Mandushev_Torres_Latham:apj:2005a}.
Instead, we do not detect any variation exceeding the measurement uncertainties (see Figure~\ref{fig:rvtres3}). 
We conclude that the RV variations are real, and the star is indeed orbited by a Jovian
planet.

\section{Estimates of Planet Parameters and Conclusions} 
\label{sec:discuss}

We analyze our photometric time series using the analytic transit curves of \cite{Mandel_Agol:apjl:2002a} and the Markov Chain Monte Carlo (MCMC) techniques described in \cite{Holman_Winn_Latham:apj:2006a}, \cite{Charbonneau_Winn_Everett:apj:2007a}, and \cite{Winn_Holman_Roussanova:apj:2007a}.
We assume a circular orbit of constant $P$.
We first estimate $T_c$ by fitting a model light curve (as described below) to only the $z$ data. 
We then determine $P$ by phase-folding the $z$ data with the TrES and HAT discovery data (which affords a baseline of 1.8\,years) while varying the trial values for $P$. 
We then fix the values for $T_c$ and $P$ (stated in Table~\ref{tab:tres3b}) in the subsequent analysis.

The values of the planet radius $R_p$ and the stellar radius $R_{\star}$ as constrained by the light curves are covariant with $M_{\star}$. 
In our MCMC analysis, we fix $M_{\star} = 0.9\,M_{\Sun}$, and then estimate the systematic error in the radii using the scaling relations $R_p \propto R_{\star} \propto M_{\star}^{1/3}$ (see footnote to Table~\ref{tab:tres3}). 
We assume a quadratic limb-darkening law with coefficients fixed at the band-dependent values tabulated in \cite{Claret:aa:2000a, Claret:aa:2004a} for the spectroscopically-estimated $T_{\rm eff}$ and $\log{g}$ and assuming solar metallicity.

\begin{figure}
\epsscale{0.8}
\plotone{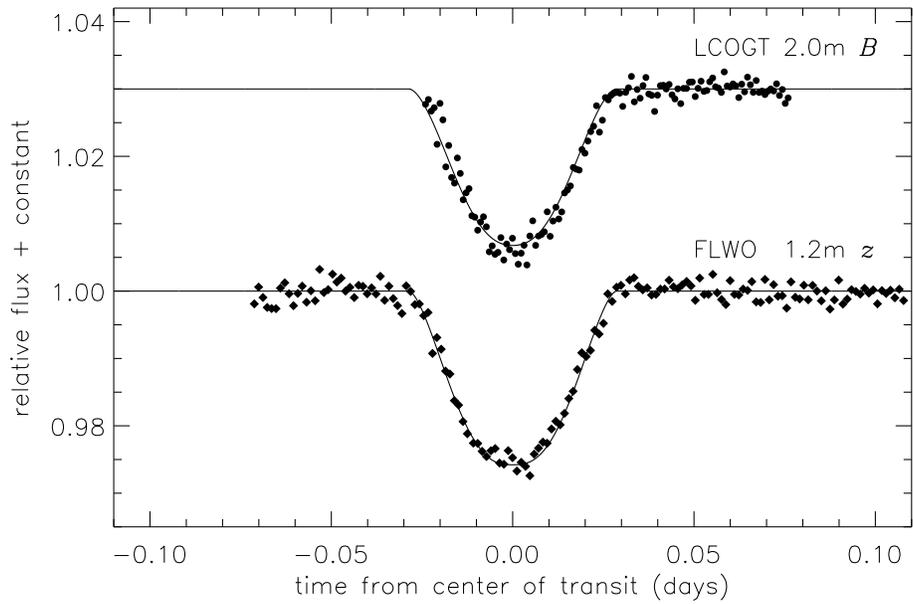}
\caption{Relative flux of the TrES-3 system as a function of time from the center of transit, adopting the ephemeris in Table~\ref{tab:tres3b}, and including the residual color-dependent extinction correction (\S\ref{sec:discuss}). 
Each of these follow-up light curves is labeled with the telescope and filter employed. 
We have overplotted the simultaneous best-fit solution, adopting the appropriate quadratic limb-darkening parameters for each band pass.%
}
\label{fig:treslc}
\end{figure}

The remaining free parameters in our model are $R_p$, $R_{\star}$, and $i$. 
We require two additional parameters, $k_B$ and $k_z$, the respective residual color-dependent extinction to the $B$ and $z$ light curves, assuming that the observed flux is proportional to $\exp{(-k\, m)}$, where $m$ denotes the airmass. 
We find that the TrES and HAT discovery data are too noisy to meaningfully constrain the parameters, and so we restrict our analysis to the $B$ and $z$ data. 
We first find the values of $R_p$, $R_{\star}$, $i$, $k_B$, and $k_z$ that minimize ${\chi}^2$ using the AMOEBA algorithm \citep[][]{Press_Teukolsky_Vetterling:1992a}.
This model is shown as the solid curves in Figure~\ref{fig:treslc}. 
We then created 2 MCMC chains with $428,\!000$ points each, one starting from the best-fit values and one starting a randomly-generated perturbation to these values. 
We subsequently rejected the first $100,\!000$ points to minimize the effect of the initial conditions, and found the results of the two chains to be indistinguishable.
We then examined the histograms of the 5 input parameter values, as well as the histograms for several combinations of parameters relevant to anticipated follow-up studies. 
We assigned the optimal value to be the median, and the 1-$\sigma$ error to be the symmetric range about the median that encompassed 68.3\% of the values. 
We state our estimates of the parameters in Table~\ref{tab:tres3b}.
The estimated values for $k_B$ ($0.0029 \pm 0.0006$) and $k_z$ ($-0.0021 \pm 0.0004$) are small and of opposite sign, which is consistent with a modest difference between the average color of the calibration field stars and the target.

Despite the V-shaped transit, our best-fit values indicate that  \tresThreeS\ is not grazing, i.e.\ the disk of the planet is entirely contained within the disk of the star at mid-transit, although grazing solutions are permitted by the data. 
Importantly, our ability to obtain well-constrained estimates of $R_{p}$ and $R_{\star}$ despite the large $b$ hinged on having observations of the transit in both $B$ and $z$. 
The large difference in central wavelength and hence stellar limb-darkening between these two bands permitted us to rule out a family of degenerate solutions that is allowed by observations in only a single color. 
We tested this notion by repeating the analysis above for only the $z$ data, and found that values of $R_p$ as large as $2.0\,R_{\rm Jup}$ could not be excluded.

\tresThreeP\ presents a useful testbed for theoretical models of gas giants.
Its radius places it in the growing family of planets with radii that exceed that predicted for models of irradiated gas giants. 
It is one of the most massive transiting planets, and has one of the shortest periods.
We recall that the discovery of \ogletrFSb\ at a distance of only 0.023\,AU from its star stimulated investigations into the timescales for orbital decay and thermal evaporation. 
The comparable orbital separation of \tresThreeP\ implies that many of these estimates are directly applicable to the new planet, but with the important difference that the much brighter apparent magnitude affords the opportunity for more precise study.
In particular, we note that direct searches for decay of the orbital period may inform our understanding of dissipation in stellar convective zones \citep{Sasselov:apj:2003a}, particularly since both the \tresThreeP\ planet and its stellar convective zone are more massive than that of \ogletrFSb.
Furthermore, the mass and orbital separation of \tresThreeP\ are intriguingly close to the critical values estimated by \cite{Baraffe_Selsis_Chabrier:aa:2004a} below which evaporation would become a dominant process.
The measurement of the angle between the planetary orbit and stellar spin axis of \tresThreeS\ may detect the substantial misalignment that might be expected for planets that were tidally captured rather than migrating inwards \citep{Gaudi_Winn:apj:2007a}.
Assuming isotropic emission, the equilibrium temperature of \tresThreeP\ is $1643(1-A)^{1/4}\,K $, where $A$ is the Bond albedo. 
We intend to obtain {\it Spitzer} observations of \tresThreeP, as we have previously done for \tresOne\ and \tresTwo.
Finally, we note that \tresThreeP\ is extremely favorable for attempts to detect reflected starlight \citep{Charbonneau_Noyes_Korzennik:apjl:1999a, Leigh_Collier-Cameron_Udry:mnras:2003a, Rowe_Matthews_Seager:apj:2006a} and thus determine the geometric albedo, $p_{\lambda}$ of the planet.
The flux of the planet near opposition relative to that of the star is $p_{\lambda} \times (R_p / a)^2 = p_{\lambda} \times 7.5 \times 10^{-4}$, which is more than twice that of any of the other known nearby transiting planets.

\acknowledgments
We thank B.~S.~Gaudi for a useful discussion.
We thank the referee for helpful comments that improved the paper. 
This material is based upon work supported by NASA under grants NNG04GN74G, NNG04LG89G, NNG05GI57G, NNG05GJ29G, and NNH05AB88I issued through the Origins of Solar Systems Program. 
We acknowledge support from the NASA \textit{Kepler} mission under Cooperative Agreement NCC2-1390.
Work by G.~\'{A}.~B.\ was supported by NASA through Hubble Fellowship Grant HST-HF-01170.01-A.
G.~K.\ acknowledges the support of OTKA~K-60750.

\clearpage

\begin{deluxetable}{lcc}
\tablewidth{0pt}
\tablecaption{TrES-3 Parent Star \label{tab:tres3}}
\tablehead{ \colhead{Parameter} & \colhead{Value}  &  \colhead{Reference} }
\startdata
R.A. \phm{00000000.} (J2000)  &  \phm{0}$17^{\rm h} 52^{\rm m} 07\fs03$ &  \\
Decl. \phm{00000000} (J2000)  &  $+37\arcdeg 32\arcmin 46\farcs1$ &  \\
GSC & \mbox{\object[GSC 03089-00929]{03089-00929}} & \\
$M_{\star}$ \phm{0000000000} ($M_{\sun}$)  &  \phm{.}$0.90\pm0.15$  & 1 \\ 
$R_{\star}$ \tablenotemark{a} \phm{000000000} ($R_{\sun}$) &  \phm{0}$0.802 \pm 0.046$  & 1 \\ 
$T_{\rm eff}$ \phm{0000000000} (K) & $5720 \pm 150$ & 1 \\ 
$\log{g}$ \phm{000000000} (dex) & \phm{0}$4.6 \pm 0.3$ & 1 \\ 
$v\sin{i}$ \phm{00000000} (${\rm km\, s^{-1}}$) & $< 2$ & 1 \\ 
$V$ \phm{00000000000.} (mag) & $12.402\pm0.006$ & 1 \\ 
$B-V$ \phm{0000000.} (mag) &  \phn$0.712\pm0.009$ & 1 \\
$V-R_{\rm C}$  \phm{$000000.$} (mag) &  \phn$0.417\pm0.010$ & 1\\
$V-I_{\rm C}$  \phm{$0000000$} (mag) &  \phn$0.799\pm0.010$ & 1\\
$J$  \phm{000000000000}  (mag) &  $11.015 \pm 0.022$ & 2 \\
$J-H$  \phm{0000000.} (mag) & \phn$0.360 \pm 0.030$ & 2 \\
$J-K_{s}$  \phm{0000000} (mag) & \phn$0.407 \pm 0.028$ & 2 \\
$[\mu_{\alpha},\mu_{\delta}]$ \phm{0000000} ($\mathrm{mas\ yr^{-1}}$) &  $[-22.5,32.0]$ & 3 \\
\enddata
\tablerefs{%
(1) This work; 
(2) 2MASS Catalog;
(3) UCAC2 Bright Star Supplement%
}
\tablenotetext{a}{%
The uncertainty in $R_{\star}$ includes both the statistical error and the 5.6\% uncertainty resulting from uncertainty in $M_{\star}$.%
}
\end{deluxetable}
 
\clearpage

\begin{deluxetable}{lcc}
\tablewidth{0pt}
\tablecaption{Relative radial-velocity measurements of TrES-3 \label{tab:rvtres3}}
\tablehead{ \colhead{Observation Epoch} & \colhead{Radial Velocity}  &  \colhead{$\sigma_{\rm RV}$} \\
 \colhead{HJD - $2,\!400,\!000$} & \colhead{m s$^{-1}$}  &  \colhead{m s$^{-1}$}}
\startdata
54186.99525 &   \phm{.}229.0 &  10.8 \\
54187.11354 &    \phm{$.0$}82.1 &    \phm{$0$}8.5 \\
54187.96330 &   \phm{.}167.5 &   \phm{$0$}8.4 \\
54188.04216 &   \phm{.}286.3 &   \phm{$0$}9.7 \\
54188.96503 & -336.5 &   \phm{$0$}8.6 \\
54189.04672 & -234.5 &  10.9 \\
54189.09622 & -183.9 &   \phm{$0$}9.3 \\
\enddata
\end{deluxetable}
 
\clearpage

\begin{deluxetable}{lc}
\tablewidth{0pt}
\tablecaption{TrES-3 Planet \label{tab:tres3b}}
\tablehead{ \colhead{Parameter} & \colhead{Value} }
\startdata
$P$ \phm{000.} (d) &  \phm{000000} $1.30619\pm 0.00001$ \\ 
$T_{c}$ \phm{000} (HJD)  & \phm{.}$2454185.9101\pm 0.0003$ \\ 
$a/R_{\star}$ & \phm{000000.}$6.06 \pm 0.10$ \\ 
$a$ \phm{0000} (AU) &  \phm{000000} $0.0226\pm0.0013$ \\ 
$b=a \cos{i} / R_{\star}$ & \phm{000000.}$0.8277 \pm 0.0097$ \\ 
$i$ \phm{0000.} ($\degr$)  &  \phm{00000} $82.15\pm0.21$  \\ 
$K$ \phm{0000}($\mathrm{m\,s^{-1}}$) &  \phm{0000} $378.4\pm9.9$ \\ 
$\gamma$ \phm{0000.}($\mathrm{m\,s^{-1}}$) &  \phm{000.} $-70.6\pm6.2$ \\ 
$M_{p}$  \phm{00.} ($M_{\rm Jup}$) &  \phm{000000} $1.92 \pm 0.23$ \\ 
$R_{p}/R_{\star}$ &  \phm{000000} $0.1660 \pm 0.0024$ \\ 
$R_{p}$ \tablenotemark{a}  \phm{00.}($R_{\rm Jup}$ \tablenotemark{b}) &  \phm{000000} $1.295 \pm 0.081$ \\ 
\enddata
\tablenotetext{a}{%
The uncertainty in $R_p$ includes both the statistical error and the 5.6\% uncertainty resulting from uncertainty in $M_{\star}$ (Table~\ref{tab:tres3}).%
}
\tablenotetext{b}{%
$R_{\rm Jup} = 71,\!492$~km, the equatorial radius of Jupiter at 1~bar.%
}
\end{deluxetable}
 
\clearpage

\clearpage

\begin{thebibliography}{}

\bibitem[\protect\citeauthoryear{{Bakos} et~al.}{{Bakos}
  et~al.}{2004}]{Bakos_Noyes_Kovacs:pasp:2004a}
{Bakos}, G., {Noyes}, R.~W., {Kov{\'a}cs}, G., {Stanek}, K.~Z., {Sasselov},
  D.~D.,  \& {Domsa}, I. 2004, PASP, 116, 266

\bibitem[\protect\citeauthoryear{{Bakos} et~al.}{{Bakos}
  et~al.}{2007}]{Bakos_Noyes_Kovacs:apj:2007a}
{Bakos}, G.~{\'A}., et~al. 2007, ApJ, 656, 552

\bibitem[\protect\citeauthoryear{{Baraffe} et~al.}{{Baraffe}
  et~al.}{2004}]{Baraffe_Selsis_Chabrier:aa:2004a}
{Baraffe}, I., {Selsis}, F., {Chabrier}, G., {Barman}, T.~S., {Allard}, F.,
  {Hauschildt}, P.~H.,  \& {Lammer}, H. 2004, A\&A, 419, L13

\bibitem[\protect\citeauthoryear{{Bouchy} et~al.}{{Bouchy}
  et~al.}{2004}]{Bouchy_Pont_Santos:aa:2004a}
{Bouchy}, F., {Pont}, F., {Santos}, N.~C., {Melo}, C., {Mayor}, M., {Queloz},
  D.,  \& {Udry}, S. 2004, A\&A, 421, L13

\bibitem[\protect\citeauthoryear{{Bouchy} et~al.}{{Bouchy}
  et~al.}{2005}]{Bouchy_Udry_Mayor:aa:2005a}
{Bouchy}, F., et~al. 2005, A\&A, 444, L15

\bibitem[\protect\citeauthoryear{{Charbonneau} et~al.}{{Charbonneau}
  et~al.}{1999}]{Charbonneau_Noyes_Korzennik:apjl:1999a}
{Charbonneau}, D., {Noyes}, R.~W., {Korzennik}, S.~G., {Nisenson}, P., {Jha},
  S., {Vogt}, S.~S.,  \& {Kibrick}, R.~I. 1999, ApJ, 522, L145

\bibitem[\protect\citeauthoryear{{Charbonneau} et~al.}{{Charbonneau}
  et~al.}{2007}]{Charbonneau_Winn_Everett:apj:2007a}
{Charbonneau}, D., {Winn}, J.~N., {Everett}, M.~E., {Latham}, D.~W., {Holman},
  M.~J., {Esquerdo}, G.~A.,  \& {O'Donovan}, F.~T. 2007, ApJ, 658, 1322

\bibitem[\protect\citeauthoryear{{Claret}}{{Claret}}{2000}]{Claret:aa:2000a}
{Claret}, A. 2000, A\&A, 363, 1081

\bibitem[\protect\citeauthoryear{{Claret}}{{Claret}}{2004}]{Claret:aa:2004a}
{Claret}, A. 2004, A\&A, 428, 1001

\bibitem[\protect\citeauthoryear{{Collier Cameron} et~al.}{{Collier Cameron}
  et~al.}{2007}]{Collier-Cameron_Bouchy_Hebrard:MNRAS:2007a}
{Collier Cameron}, A., et~al. 2007, MNRAS, 375, 951

\bibitem[\protect\citeauthoryear{{Dunham} et~al.}{{Dunham}
  et~al.}{2004}]{Dunham_Mandushev_Taylor:pasp:2004a}
{Dunham}, E.~W., {Mandushev}, G.~I., {Taylor}, B.~W.,  \& {Oetiker}, B. 2004,
  PASP, 116, 1072

\bibitem[\protect\citeauthoryear{{Fressin} et~al.}{{Fressin}
  et~al.}{2007}]{Fressin_Guillot_Morello:preprint:2007a}
{Fressin}, F., {Guillot}, T., {Morello}, V.,  \& {Pont}, F. 2007, A\&A, in
  press (arXiv:0704.1919), 704

\bibitem[\protect\citeauthoryear{{Gaudi}, {Seager}, \&
  {Mallen-Ornelas}}{{Gaudi}
  et~al.}{2005}]{Gaudi_Seager_Mallen-Ornelas:apj:2005a}
{Gaudi}, B.~S., {Seager}, S.,  \& {Mallen-Ornelas}, G. 2005, ApJ, 623, 472

\bibitem[\protect\citeauthoryear{{Gaudi} \& {Winn}}{{Gaudi} \&
  {Winn}}{2007}]{Gaudi_Winn:apj:2007a}
{Gaudi}, B.~S.,  \& {Winn}, J.~N. 2007, ApJ, 655, 550

\bibitem[\protect\citeauthoryear{{Gould} et~al.}{{Gould}
  et~al.}{2006}]{Gould_Dorsher_Gaudi:acta:2006a}
{Gould}, A., {Dorsher}, S., {Gaudi}, B.~S.,  \& {Udalski}, A. 2006, Acta
  Astron., 56, 1

\bibitem[\protect\citeauthoryear{{Holman} et~al.}{{Holman}
  et~al.}{2006}]{Holman_Winn_Latham:apj:2006a}
{Holman}, M.~J., et~al. 2006, ApJ, 652, 1715

\bibitem[\protect\citeauthoryear{{Konacki} et~al.}{{Konacki}
  et~al.}{2003}]{Konacki_Torres_Jha:nat:2003a}
{Konacki}, M., {Torres}, G., {Jha}, S.,  \& {Sasselov}, D.~D. 2003, Nature,
  421, 507

\bibitem[\protect\citeauthoryear{{Konacki} et~al.}{{Konacki}
  et~al.}{2004}]{Konacki_Torres_Sasselov:apjl:2004a}
{Konacki}, M., et~al. 2004, ApJ, 609, L37

\bibitem[\protect\citeauthoryear{{Kov{\'a}cs}, {Bakos}, \&
  {Noyes}}{{Kov{\'a}cs} et~al.}{2005}]{Kovacs_Bakos_Noyes:mnras:2005a}
{Kov{\'a}cs}, G., {Bakos}, G.,  \& {Noyes}, R.~W. 2005, MNRAS, 356, 557

\bibitem[\protect\citeauthoryear{{Kov{\'a}cs}, {Zucker}, \&
  {Mazeh}}{{Kov{\'a}cs} et~al.}{2002}]{Kovacs_Zucker_Mazeh:aa:2002a}
{Kov{\'a}cs}, G., {Zucker}, S.,  \& {Mazeh}, T. 2002, A\&A, 391, 369

\bibitem[\protect\citeauthoryear{{Landolt}}{{Landolt}}{1992}]{Landolt:aj:1992a}
{Landolt}, A.~U. 1992, AJ, 104, 340

\bibitem[\protect\citeauthoryear{{Latham}}{{Latham}}{1992}]{Latham:ASP:1992a}
{Latham}, D.~W. 1992, in ASP Conf. Ser. 32, ed. H.~A. {McAlister} \& W.~I.
  {Hartkopf} (San Francisco: ASP), 110

\bibitem[\protect\citeauthoryear{{Leigh} et~al.}{{Leigh}
  et~al.}{2003}]{Leigh_Collier-Cameron_Udry:mnras:2003a}
{Leigh}, C., {Collier Cameron}, A., {Udry}, S., {Donati}, J.-F., {Horne}, K.,
  {James}, D.,  \& {Penny}, A. 2003, MNRAS, 346, L16

\bibitem[\protect\citeauthoryear{{Mandel} \& {Agol}}{{Mandel} \&
  {Agol}}{2002}]{Mandel_Agol:apjl:2002a}
{Mandel}, K.,  \& {Agol}, E. 2002, ApJ, 580, L171

\bibitem[\protect\citeauthoryear{{Mandushev} et~al.}{{Mandushev}
  et~al.}{2005}]{Mandushev_Torres_Latham:apj:2005a}
{Mandushev}, G., et~al. 2005, ApJ, 621, 1061

\bibitem[\protect\citeauthoryear{{Mazeh}, {Zucker}, \& {Pont}}{{Mazeh}
  et~al.}{2005}]{Mazeh_Zucker_Pont:mnras:2005a}
{Mazeh}, T., {Zucker}, S.,  \& {Pont}, F. 2005, MNRAS, 356, 955

\bibitem[\protect\citeauthoryear{{O'Donovan} et~al.}{{O'Donovan}
  et~al.}{2007}]{ODonovan_Charbonneau_Alonso:preprint:2007a}
{O'Donovan}, F.~T., et~al. 2007, ApJ, accepted (astro-ph/0610603)

\bibitem[\protect\citeauthoryear{{O'Donovan} et~al.}{{O'Donovan}
  et~al.}{2006a}]{ODonovan_Charbonneau_Mandushev:apjl:2006a}
{O'Donovan}, F.~T., et~al. 2006a, ApJ, 651, L61

\bibitem[\protect\citeauthoryear{{O'Donovan} et~al.}{{O'Donovan}
  et~al.}{2006b}]{ODonovan_Charbonneau_Torres:apj:2006a}
{O'Donovan}, F.~T., et~al. 2006b, ApJ, 644, 1237

\bibitem[\protect\citeauthoryear{{Pont} et~al.}{{Pont}
  et~al.}{2005}]{Pont_Bouchy_Melo:aa:2005a}
{Pont}, F., {Bouchy}, F., {Melo}, C., {Santos}, N.~C., {Mayor}, M., {Queloz},
  D.,  \& {Udry}, S. 2005, A\&A, 438, 1123

\bibitem[\protect\citeauthoryear{{Press} et~al.}{{Press}
  et~al.}{1992}]{Press_Teukolsky_Vetterling:1992a}
{Press}, W.~H., {Teukolsky}, S.~A., {Vetterling}, W.~T.,  \& {Flannery}, B.~P.
  1992, {Numerical Recipes in C} (Cambridge: Cambridge University Press)

\bibitem[\protect\citeauthoryear{{Queloz} et~al.}{{Queloz}
  et~al.}{2001}]{Queloz_Henry_Sivan:aa:2001a}
{Queloz}, D., et~al. 2001, A\&A, 379, 279

\bibitem[\protect\citeauthoryear{{Rowe} et~al.}{{Rowe}
  et~al.}{2006}]{Rowe_Matthews_Seager:apj:2006a}
{Rowe}, J.~F., et~al. 2006, ApJ, 646, 1241

\bibitem[\protect\citeauthoryear{{Saar}, {Butler}, \& {Marcy}}{{Saar}
  et~al.}{1998}]{Saar_Butler_Marcy:apjl:1998a}
{Saar}, S.~H., {Butler}, R.~P.,  \& {Marcy}, G.~W. 1998, ApJ, 498, L153

\bibitem[\protect\citeauthoryear{{Santos} et~al.}{{Santos}
  et~al.}{2000}]{Santos_Mayor_Naef:aa:2000a}
{Santos}, N.~C., {Mayor}, M., {Naef}, D., {Pepe}, F., {Queloz}, D., {Udry}, S.,
   \& {Blecha}, A. 2000, A\&A, 361, 265

\bibitem[\protect\citeauthoryear{{Santos} et~al.}{{Santos}
  et~al.}{2002}]{Santos_Mayor_Naef:aa:2002a}
{Santos}, N.~C., et~al. 2002, A\&A, 392, 215

\bibitem[\protect\citeauthoryear{{Sasselov}}{{Sasselov}}{2003}]{Sasselov:apj:2%
003a}
{Sasselov}, D.~D. 2003, ApJ, 596, 1327

\bibitem[\protect\citeauthoryear{{Sozzetti} et~al.}{{Sozzetti}
  et~al.}{2007}]{Sozzetti_Torres_Charbonneau:preprint:2007a}
{Sozzetti}, A., {Torres}, G., {Charbonneau}, D., {Latham}, D.~W., {Holman},
  M.~J., {Winn}, J.~N., {Laird}, J.~B.,  \& {O'Donovan}, F.~T. 2007, ApJ, in
  press (arXiv:0704.2938), 704

\bibitem[\protect\citeauthoryear{{Sozzetti} et~al.}{{Sozzetti}
  et~al.}{2006}]{Sozzetti_Torres_Latham:apj:2006a}
{Sozzetti}, A., {Torres}, G., {Latham}, D.~W., {Carney}, B.~W., {Stefanik},
  R.~P., {Boss}, A.~P., {Laird}, J.~B.,  \& {Korzennik}, S.~G. 2006, ApJ, 649,
  428

\bibitem[\protect\citeauthoryear{{Sozzetti} et~al.}{{Sozzetti}
  et~al.}{2004}]{Sozzetti_Yong_Torres:apjl:2004a}
{Sozzetti}, A., et~al. 2004, ApJ, 616, L167

\bibitem[\protect\citeauthoryear{{Torres} et~al.}{{Torres}
  et~al.}{2005}]{Torres_Konacki_Sasselov:apj:2005a}
{Torres}, G., {Konacki}, M., {Sasselov}, D.~D.,  \& {Jha}, S. 2005, ApJ, 619,
  558

\bibitem[\protect\citeauthoryear{{Trilling} et~al.}{{Trilling}
  et~al.}{1998}]{Trilling_Benz_Guillot:apj:1998a}
{Trilling}, D.~E., {Benz}, W., {Guillot}, T., {Lunine}, J.~I., {Hubbard},
  W.~B.,  \& {Burrows}, A. 1998, ApJ, 500, 428

\bibitem[\protect\citeauthoryear{{Udalski} et~al.}{{Udalski}
  et~al.}{2003}]{Udalski_Pietrzynski_Szymanski:acta:2003a}
{Udalski}, A., {Pietrzynski}, G., {Szymanski}, M., {Kubiak}, M., {Zebrun}, K.,
  {Soszynski}, I., {Szewczyk}, O.,  \& {Wyrzykowski}, L. 2003, Acta Astron.,
  53, 133

\bibitem[\protect\citeauthoryear{{Vogt} et~al.}{{Vogt}
  et~al.}{1994}]{Vogt_Allen_Bigelow:SPIE:1994a}
{Vogt}, S.~S., et~al. 1994, in Instrumentation in Astronomy VIII, D. L.
  Crawford \& E. R. Craine eds., Proc. SPIE, 2198, 362

\bibitem[\protect\citeauthoryear{{Winn}, {Holman}, \& {Roussanova}}{{Winn}
  et~al.}{2007}]{Winn_Holman_Roussanova:apj:2007a}
{Winn}, J.~N., {Holman}, M.~J.,  \& {Roussanova}, A. 2007, ApJ, 657, 1098

\bibitem[\protect\citeauthoryear{{Wright}}{{Wright}}{2005}]{Wright:pasp:2005a}
{Wright}, J.~T. 2005, \pasp, 117, 657

\end{thebibliography}
\end{document}